\documentclass[aps,prb,showpacs,twocolumn,showkeys,amssymb,amsmath,superscriptaddress,reprint]{revtex4-1}

\usepackage{lipsum}
\usepackage{amsmath}
\usepackage{amsfonts}
\usepackage{amssymb}
\usepackage{graphicx}
\usepackage{color}
\usepackage{verbatim}

\newcommand{\mtx}[1]{\ensuremath{\mathbf{#1}}}
\DeclareMathOperator{\Tr}{Tr}

\begin{document}
\title{Landauer Current and Mutual Information}
\author{Auditya Sharma}
\affiliation{School of Chemistry, The Sackler Faculty of Exact
  Sciences, Tel Aviv University, Tel Aviv 69978, Israel}

\author{Eran Rabani}
\affiliation{Department of Chemistry, University of California and
  Lawrence Berkeley National Laboratory, Berkeley, California 94720,
  USA}
\begin{abstract}
We study quantum evolution of the mutual information of a quantum dot
connected to left and right leads initially maintained at chemical
potentials $\mu_{L}$ and $\mu_{R}$ respectively, within the
non-interacting resonant-level model. The full nonequilirbium mixed
state density matrix of the whole system is written down exactly, and
the mutual information of the dot with respect to the leads is computed. 
A strong and direct correlation is found between the
Landauer current, and the mutual information at all times, the steady-state
values in particular displaying a quadratic relationship at high
temperatures. Strikingly, it is found that one can obtain a maximal MI
 by simply applying a sufficiently large
`source-drain' voltage $V_{SD}$ even at high temperatures.
\end{abstract}

\maketitle

\section{Introduction:}
The tunneling current across a quantum dot in a non-interacting
electron framework has been studied for some five decades
now~\cite{haug2007quantum,datta1997electronic,landauer1957spatial,landauer1970electrical,buttiker1986four},
and is widely regarded as well-understood because it is amenable to an
analytical solution within the powerful non-equilibrium Green function
framework~\cite{haug2007quantum,stefanucci_nonequilibrium_2013}. Recently
developed numerical
methods~\cite{lothar2008real,weiss_iterative_2008,werner_diagrammatic_2009,Wang2009,gull2010bold,Segal10,Huetzen12,Simine13}
have made the study of interacting systems driven away from
equilibrium an accessible task as well.  Although the notion of
entanglement, arguably the most fundamental aspect of `quantum
physics' has been around since the birth of quantum mechanics, it is
only in the last decade or so that its study in various contexts of
physics has become
popular~\cite{amico2008entanglement,znidaric2012,eisler2014,znidaric2008}. Understandably, this has
coincided with the development of a whole machinery of mathematical
tools to quantify it~\cite{plenio2005introduction}, and simultaneous
advances in numerical approaches~\cite{schollwock2005density} which
allow for a direct utilization of the machinery. It is of great
current interest in almost every discipline of physics to explore
hitherto unexpected connections to entanglement.

Here, we investigate the connection between mutual information and Landauer
current in a simple non-interacting resonant-level model. Related
recent studies have either restricted themselves to looking at the
system as a pure state in order to extract an exact useful
relationship with quantum
noise~\cite{klich2009quantum,chien2014landauer} or have briefly hinted
the possibility of calculating entanglement as an
aside~\cite{dhar2012nonequilibrium}. Here, with mutual information of the
quantum dot with respect to the leads for the full nonequilibrium
mixed state as our focus, we derive explicitly how the nonequilibrium
density matrix of the system can be expressed as an effective thermal
density matrix. This result is in agreement with the general thermal
density matrix prespcription that was developed formally within a
so-called $Y$ operator approach a while back by
Hershfield~\cite{hershfield1993reformulation}.

We start by recalling how the non-equilibrium current is recovered
with the aid of a simple prescription.  Employing the methods of Ingo
Peschel~\cite{peschel2003letter} we then proceed to obtain the
von-Neumann entropy of any part of the full system with respect to its
bath in terms of a correlation matrix. Introducing the notion of
mutual information~\cite{henderson2001classical,wolf2008area} we are then able to calculate and study the
total correlation between the dot (which is taken as our system) and the
bath (given by the two leads). A strong correlation between the
current and the entanglment between the dot and the leads is
seen. Surprisingly, the maximal MI of the quantum dot with respect to the leads
is attained at steady-state by the application of a large
`source-drain' voltage, unlike the effect of temperature which tends
to cause the mutual information to fall, thus making voltage and temperature
not alike in this respect.

\section{The Model}
We consider the following tunneling Hamiltonian:
\begin{align}
\label{eqn:ham}
\mathcal{H} &= \mathcal{H}_{D}+\mathcal{H}_{L}+\mathcal{H}_{R}+\mathcal{H}_{LD}+\mathcal{H}_{RD}\\
\mathcal{H}_{D} &= \varepsilon_{d}d^{\dagger}d \nonumber \\
\mathcal{H}_{L(R)} &= \sum_{k\in L(R)}\varepsilon_{k}c_{k}^{\dagger}c_{k}\nonumber \\ 
\mathcal{H}_{LD(RD)} &= \sum_{k\in L(R)}t_{k}(d^{\dagger}c_{k}+c_{k}^{\dagger}d)\nonumber,
\end{align}
where $\varepsilon_{d}$ is the energy of the isolated quantum dot,
$\varepsilon_{k}$ is the energy of the electrode mode $k$, and $t_{k}$
is the coupling between the the quantum dot and the electrode mode $k$
(with the energy leves and the couplings of both left and right
electrodes assumed to be identical). The `wide-band
limit'~\cite{lothar2008real,swenson2011application} with a sharp
cutoff at high and low energy values is imposed:
\begin{equation}
J_{L/R}(\varepsilon) = \frac{\Gamma_{L/R}}{(1+e^{A(\varepsilon-\frac{B}{2})})(1+e^{-A(\varepsilon+\frac{B}{2})})},
\end{equation} 
with $\Gamma_{L}=\Gamma_{R}=\frac{1}{2}$, $\Gamma = \Gamma_{L}+\Gamma_{R}$, $A = 5\Gamma$, and $B=20\Gamma$. We set $\hbar=1$, $k_{B}=1$, and electronic charge $e=1$ throughout. With a uniform discretization the couplings are given by $t_{k}(\varepsilon_{k}) = \sqrt{\frac{J(\varepsilon_{k})\Delta\varepsilon}{2\pi}}$.
Further, we assume that at time $t=0$, the left and right leads are
separately at thermal equilibrium each at the same temperature $T$ and
characterized by chemical potentials $\mu_{L}$ and $\mu_{R}$
respectively with $\mu_{L}-\mu_{R}\equiv e V_{SD}$ being the applied
external `voltage', and the tunneling Hamiltonian between the leads is
suddenly turned on at time $t=0$. At time $t=0$, therefore we have the
full system in the following mixed but separable state:
\begin{equation}
\rho(0) = \rho_{L}(0)\otimes\rho_{R}(0)\otimes\rho_{D}(0),
\end{equation}
where the left and right leads are in thermal ensembles
\begin{align}
\rho_{L}(0) = \frac{e^{-\beta(\mathcal{H}_{L}-\mu_{L}\sum_{k\in L}c_{k}^{\dagger}c_{k})}}{Z_{L}},\\ 
\rho_{R}(0) = \frac{e^{-\beta(\mathcal{H}_{R}-\mu_{R}\sum_{k\in R}c_{k}^{\dagger}c_{k})}}{Z_{R}},
\end{align}
and the dot is characterized by an arbitrary population $n_{0}$
\begin{equation}
\rho_{D}(0) = n_{0} d^{\dagger}d + (1-n_{0})dd^{\dagger}.
\end{equation}

\section{Current}
Although the thermodynamic limit of this problem can be solved, in order to study mutual information, we consider a sufficiently large but finite system (whose convergence to thermodynamic limit we verified) with $N_{L}=128$, $N_{R}=128$ sites on the left and right leads respectively. The full hopping matrix can be written explicitly as:
\begin{align}
T = 
\begin{pmatrix}
\varepsilon_{1} & 0 & \cdots & 0 & 0 & 0 & 0 & \cdots & \cdots & t_{1}\\
0          &\varepsilon_{2} & 0 & \cdots & \cdots & \cdots  &\cdots & \cdots & 0   & t_{2}\\
\vdots & \vdots & \ddots &\vdots & \vdots &\vdots & \vdots &\vdots &\vdots&\vdots\\
\cdots & \cdots & \cdots & \varepsilon_{N_{L}} & 0 & \cdots & \cdots &\cdots &\cdots & t_{N_{L}}\\
0 & 0 & \cdots &  0 & \varepsilon_{1} & 0 & 0 &\cdots &\cdots & t_{1}\\
0 & 0 & \cdots & \cdots & 0 & \varepsilon_{2} & \cdots & \cdots &\cdots & t_{2}\\ 
0 & 0 & \cdots & \cdots & 0 & 0 & \varepsilon_{3} & \cdots &\cdots & t_{3}\\
\vdots & \vdots & \vdots & \vdots & \vdots & \vdots & \vdots & \ddots & \vdots & \vdots\\ 
0 & 0 & \cdots & 0 & 0  & \cdots & 0 & 0 & \varepsilon_{N_{R}} & t_{N_{R}}\\
t_{1} & t_{2} & \cdots & t_{N_{L}} & t_{1} & t_{2} &\cdots & \cdots & t_{N_{R}} & \varepsilon_{d}
\end{pmatrix}
\end{align}

Next, we diagonalize the hopping matrix:
\begin{equation}
\sum_{j=1}^{N}T_{ij}\psi_{\alpha}(j) = e_{\alpha}\psi_{\alpha}(i),
\end{equation}
where $N$ is the total number of sites inclusive of the left, and right leads and one dot: $N = N_{L}+N_{R}+1$.
The eigenvalues $e_{\alpha}$ are ordered such that $e_{1}\le e_2\le\cdots\le e_{N}$. By defining the fermionic operators
$a_{\alpha} = \sum_{i=1}^{N}\psi_{\alpha}(i)c_{i}$, the Hamiltonian Eq.~\ref{eqn:ham} becomes $\mathcal{H}=\sum_{\alpha=1}^{N}e_{\alpha}a^{\dagger}_{\alpha}a_{\alpha}$.
The eigenstates of the Hamiltonian can all be written in the form $|E\rangle = \prod_{\alpha=1}^{N_{p}}a_{\alpha}^{\dagger}|0\rangle$,
with $N_{p}$ particles, and the energy $E = \sum_{\alpha=1}^{N_{p}}e_{\alpha}$ is simply given as the sum of the single-particle states occupied. Since every state can either be occupied or not, the total number of distinct eigenstates is $2^{N}$.

The left(right) current is given by the change in occupancy of the left(right) electrode:
\begin{align}
I_{L(R)}(t) &= -e \frac{d}{dt}\Big\langle\sum_{i\in L(R)}c_{i}^{\dagger}c_{i}\Big\rangle\\
        &= -\frac{ei}{\hbar}\Big\langle\sum_{i\in L(R)}t_{i}(d^{\dagger}c_{i}-c_{i}^{\dagger}d)\Big\rangle.
\end{align}
The overall current defined as $I(t)=\frac{I_{L}(t)-I_{R}(t)}{2}$ can be computed as (Appendix~\ref{sec:current}):
\begin{align}
\label{eqn:current}
I(t)  &= \sum_{j k}\tilde{t}_{j}\sum_{\alpha\beta}Im[e^{it(e_{\alpha}-e_{\beta})}\psi_{\alpha}(N)\psi_{\beta}^{*}(j)\psi_{\alpha}^{*}(k)\psi_{\beta}(k)]f_{k}, 
\end{align}
where $\tilde{t}_{j} = t_{j}$ if $j\in L$, and $\tilde{t}_{j} = -t_{j}$ if $j\in R$, and where $f_{k}$ is the Fermi function $f(\varepsilon_{k}-\mu_{L})$ for modes on the left lead, the Fermi function $f(\varepsilon_{k}-\mu_{R})$ for modes on the right lead, and the occupancy $n_{0}$ for the dot mode.

\section{Nonequilibrium Density Matrix}
In the Schrodinger picture, the time evolved density matrix is given by
\begin{align}
\label{eqn:rhooft}
\rho(t) &= e^{-i \mathcal{H}t/\hbar}\rho(0)e^{i \mathcal{H}t/\hbar}.
\end{align}
Since the initial density matrix is of the form $\rho(0) = \frac{e^{-\sum_{i,j}H_{i,j}(0)c_{i}^{\dagger}c_{j}}}{Z}$, Wick's theorem holds at $t=0$, and therefore (Appendix~\ref{sec:HC}), the Wick's theorem must hold at every instant of time, i.e. every higher order correlator can be written in terms of one-particle correlators:
\begin{equation}
\langle c_{n}^{\dagger}c_{m}^{\dagger}c_{l}c_{j}\rangle = \langle c_{n}^{\dagger}c_{j}\rangle\langle c_{m}^{\dagger}c_{l}\rangle-\langle c_{n}^{\dagger}c_{l}\rangle\langle c_{m}^{\dagger}c_{j}\rangle,
\end{equation}
the expectation value of an operator $A$ being defined as usual: $\langle A\rangle(t) = \Tr(\rho(t) A)$.

But we know that the density matrix is unique, and that if it takes the form
\begin{equation}
\label{eqn:density}
\rho(t) = \frac{e^{-\sum_{i,j}H_{i,j}(t)c_{i}^{\dagger}c_{j}}}{Z},
\end{equation}
then Wick's theorem holds.
Imposing this form, one can show that (Appendix~\ref{sec:HC}) the matrix $H(t)$ is simply determined by the relation
\begin{equation}
\exp(H(t)) = \frac{-C(t)+1}{C(t)},
\end{equation}
in terms of the correlator matrix defined as
\begin{equation}
C_{ij}(t) = \Tr(\rho(t) c_{i}^{\dagger}c_{j}).
\end{equation}

\section{Mutual Information}
\begin{figure}
\includegraphics[width=0.9\columnwidth]{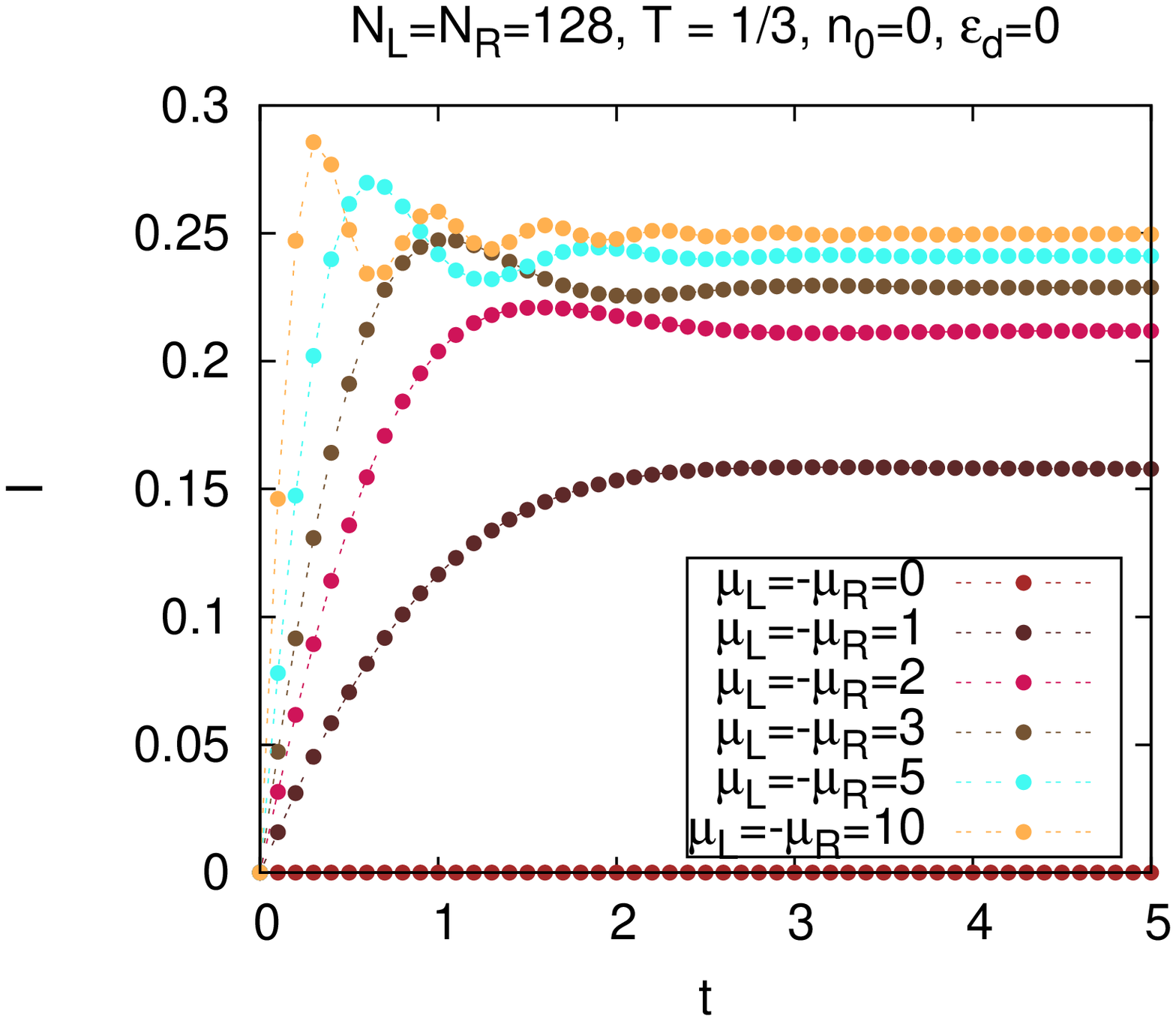}\\
\includegraphics[width=0.9\columnwidth]{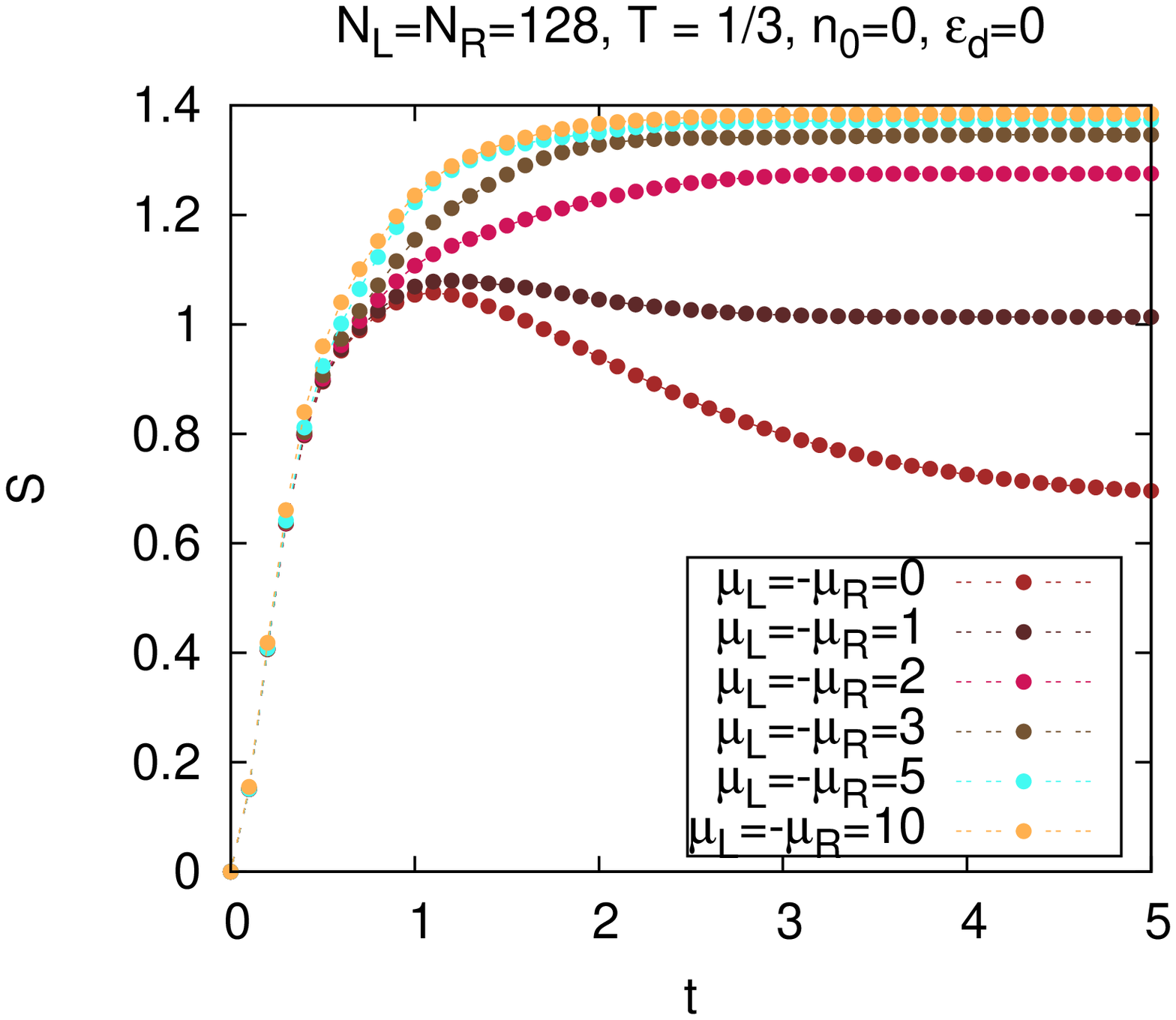}
\caption{Figure shows the intimate relationship between current and mutual information between the dot and the leads, $S$. Each of them is plotted as a function of time $t$ for just one set of parameters shown.}
\label{fig_MI_NL128}
\end{figure}

We wish to study the entanglement between the dot and the leads. The typical quantity that is used to study the entanglement between a system and its bath is the von-Neumann entropy of the reduced density matrix (of either system or bath). However in our case, this is unsuitable because the overall composite state of dot and leads given by $\rho(t)$ is a mixed state. This implies that the von-Neumann entropy of the dot would be different from the von-Neumann entropy of the leads, thus making a blind application not amenable to study bipartite entanglement. A simple workaround exists and that is to study mutual information defined as:
\begin{equation}
\label{eqn:MI}
S = S_{D}+S_{LR} - S_{full},
\end{equation}
where $S_{D} = -\Tr(\rho_{D}\ln{\rho_{D}})$, $S_{LR} = -\Tr(\rho_{LR}\ln{\rho_{LR}})$, $S_{full} = -\Tr(\rho\ln{\rho})$,
and the reduced density matrices defined as usual: $\rho_{D} = \Tr_{L,R}(\rho)$, $\rho_{LR} = \Tr_{D}(\rho)$. 

The mutual information includes both classical and quantum correlations, however it is a much studied object of interest in its own right~\cite{eisler2014,znidaric2008}.
An implicit dependence of time is assumed for all quantities above.

The reduced density matrices themselves can be written as thermal density matrices of the form $\rho_{red} = \frac{e^{-\sum_{i,j}H_{i,j}c_{i}^{\dagger}c_{j}}}{Z}$ by the same argument given in the previous section: Wick's theorem holds within the reduced subspace, and the reduced density matrix is unique, and Wick's theorem hold's for a thermal type of density matrix. Once again the $H$ matrix is related to the correlator matrix $C$ by the formula $\exp(H) = (-C+1)C^{-1}$, with the crucial difference that the correlator matrix indices run only over the reduced subspace. We now obtain a simple formula for the von-Neumann entropy of any subspace $\mathcal{G}$ 
\begin{equation}
S_{\mathcal{G}} = -\Tr(\rho_{\mathcal{G}}\ln{\rho_{\mathcal{G}}}),
\end{equation}
in terms of the eigenvalues of the correlator matrix within that subspace. With the aid of this general formula the mutual information is easily computed, by replacing $\mathcal{G}$ by $D$, $LR$, and $full$ in Eqn.~\ref{eqn:MI}. As shown in Appendix~\ref{sec:vN},
\begin{align}
\label{eqn:vN}
S_{\mathcal{G}} &= \sum_{\sigma=1}^{N_{\mathcal{G}}}\Big[-(1-C_{\sigma})\ln(1-C_{\sigma})-C_{\sigma}\ln C_{\sigma}\Big],
\end{align}
where $C_{\sigma}$ are the eigenvalues of the correlator matrix defined within the reduced subspace $\mathcal{G}$.
Thus the computation of von-Neumann entropy has been reduced from a $2^{N_{\mathcal{G}}}\times 2^{N_{\mathcal{G}}}$ eigenvalue problem to a $N_{\mathcal{G}}\times N_{\mathcal{G}}$ eigenvalue problem, where $N_{\mathcal{G}}$ is the number of sites in the subspace under consideration. This can be exploited for numerics and we do. This above formula is general and applicable for the full non-equilibrium state at arbitrary time; a similar formula was obtained for eigenstates~\cite{peschel2012special}, and for the non-equilibrium steady state~\cite{dhar2012nonequilibrium}.  

In our case, the von-Neumann entropy of the full density matrix $S_{full}$ is constant in time, as is evident from the unitary evolution of the density matrix $\rho(t)$ in Eqn.~\eqref{eqn:rhooft}, which leaves the eigenvalues of $\rho(t)$ invariant. We can write down an expression for $S_{full}$ in terms of the initial density matrix, similar to Eqn.~\eqref{eqn:vN}, where the eigenvalues $C_{\sigma}$ are just the population $n_{0}$ and Fermi functions for the left and right leads as given in Eqn.~\eqref{eqn:current}.

\section{Current and Mutual Information}
\begin{figure}
\includegraphics[width=0.9\columnwidth]{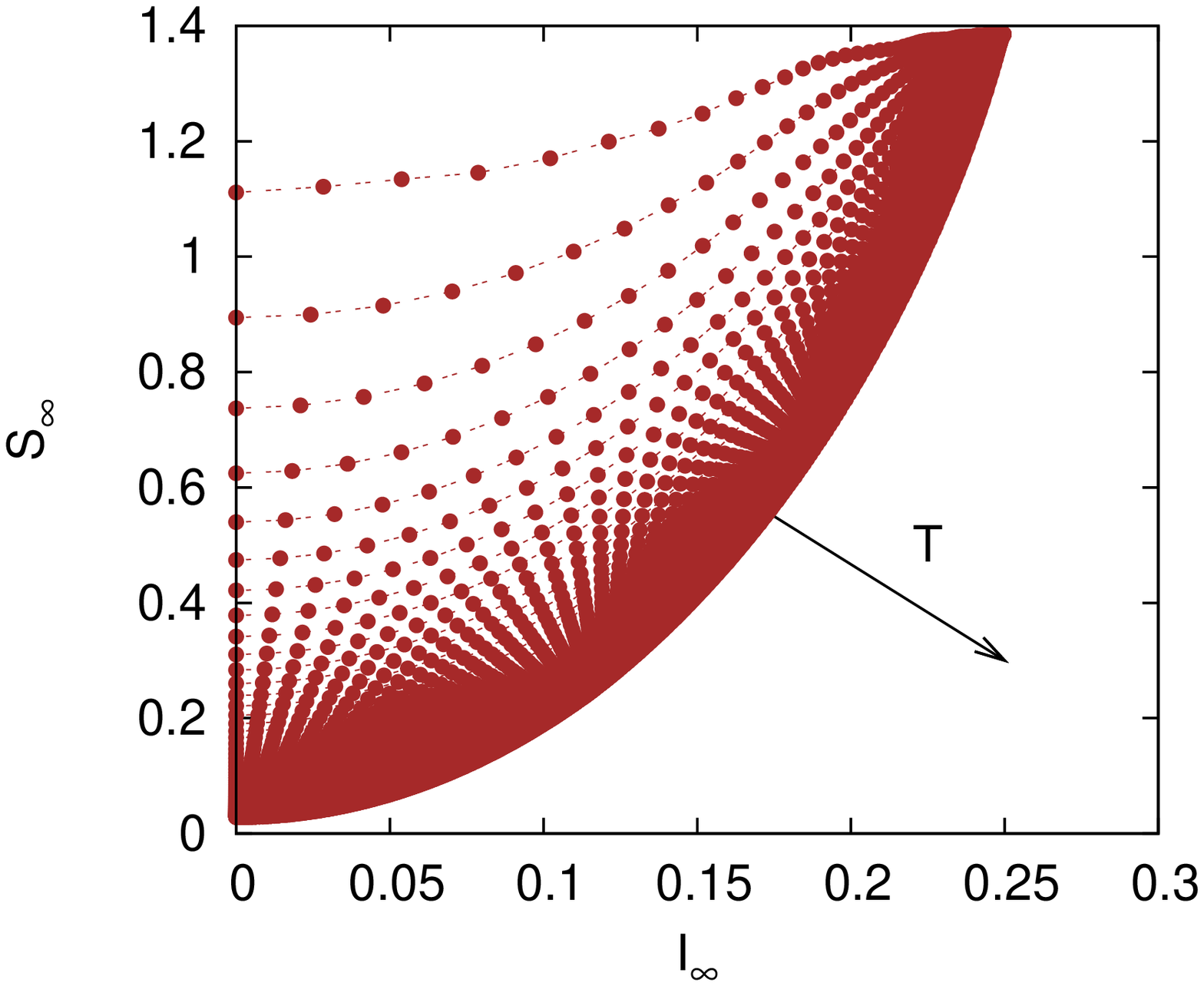}\\
\includegraphics[width=0.9\columnwidth]{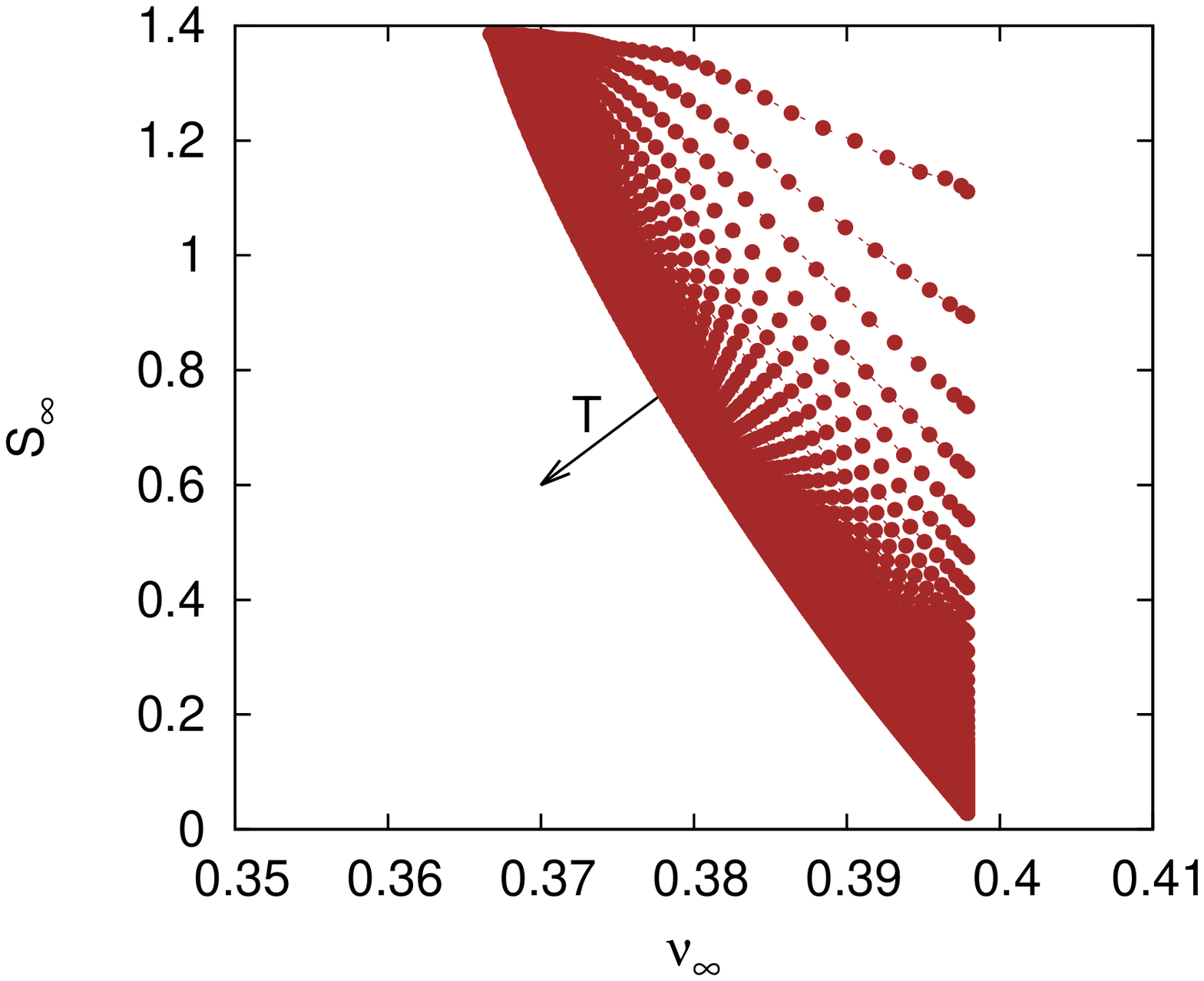}
\caption{Figure shows (a) the steady state value of $S_{\infty}$ as a function of the steady state value of current $I_{\infty}$, and (b) $S_{\infty}$ as a function of the steady-state quantum fluctuations in current $\nu_{\infty}$. A range of data are presented for temperatures going all the way from $T=0.1$ to $T=10$, the increasing direction of temperature being pointed by the arrow. At each temperature, $V_{SD}$ was varied from $0$ to $10$ and the steady state values were plotted against each other.
}
\label{fig_ss_NL128}
\end{figure}
Fig.~\ref{fig_MI_NL128} shows the intimate relation between current and mutual information, which is the heart of our finding here. We see that greater the current greater is the mutual information. In fact, it seems that the underlying mechanism by which one may attain high current is to attempt to make the dot highly entangled with its environment. Conversely, a simple method by which to obtain high MI states may be simply to maximize current in a quantum-dot-system, one that is well-studied and where considerable experimental expertise exists. We verified this direct relationship between current and MI for a series of different parameters ($T$, $\varepsilon_{d}$, $V_{SD}$) - only a representative sample is displayed in Fig.~\ref{fig_MI_NL128}. A further striking aspect of Fig.~\ref{fig_MI_NL128} is that the maximal MI of the dot ($S \sim 2\log_{e}{2}$) can be obtained even at high temperatures by the application of a large `source-drain' voltage ($e V_{SD} \gg k_{B}T$)- an observation that could be exploited experimentally for producing maximal MI quantum dots.
In order to see if further order exists between current and MI, we studied the steady state dependence of each other for a range of parameters. We found that at sufficiently high-temperatures, $S_{\infty}$ clearly goes quadratically as a function of $I_{\infty}$. Furthermore, we studied the quantum fluctuations in current at steady state defined by:
\begin{equation}
\nu_{\infty} = \langle\hat{I}(t_{\infty})^{2}\rangle - I(t_{\infty})^{2},
\end{equation}
which can again be computed by the methods described above (Appendix~\ref{sec:N}). 
As shown in Fig.~\ref{fig_ss_NL128}, at high temperatures we see a quadratic dependence between $S_{\infty}$ and $I_{\infty}$, and simultaneously, a linear relationship between $S_{\infty}$ and $\nu_{\infty}$ at least in the large $\nu_{\infty}$ regime. An exact infinite series expression for the steady state mutual information can be written down:
\begin{equation}
\label{eqn:Sinfty}
S_{\infty} = \sum_{k=1}^{\infty}\frac{2^{2k}}{(2k)(2k-1)}\big[\Tr(C_{full}-\frac{1}{2})^{2k}-\Tr(C_{bath}-\frac{1}{2})^{2k}\big].
\end{equation}
It is known that $I_{\infty}$ in the limit of high temperature has a linear dependence on $(\mu_{L}-\mu_{R})$ (`Ohm's law'), thus Fig.~\ref{fig_ss_NL128} implies that at high temperature $S_{\infty}$ has a quadratic depence on voltage. Our numerical checks show that this quadratic dependence on voltage appears to be true term-by-term for all orders in Eqn.~\ref{eqn:Sinfty}, an abstract implication of which is that the spectral function of the correlator matrix is quadratic. This is a remarkable, strange and non-trivial finding, and appears to be not accessible to simple arguments.
\section{Conclusions and Discussion}
We studied the dynamics of a quantum dot within the non-interacting resonant level model. With Heisenberg time-evolution we were able to obtain the exact nonequilibrium density matrix at arbitrary time, which in turn allowed us to study the Landauer current. Working with the notion of mutual information, we obtained a convenient expression by which one can study the nonequilibrium MI between the dot and the leads as a function of time. We found that the current through the system and the MI between the quantum dot and the leads, are intimately and directly related. Studying the steady-state mutual information $S_{\infty}$, we noticed that this quantity has a strong quadratic dependence with respect to the steady-state current $I_{\infty}$, particularly at high-temperatures - this appears to be a strange, non-trivial finding. We also observed that at sufficiently high-temperatures, one can find a window of where the MI is linear with respect to the quantum fluctuations in steady state current, $\nu_{\infty}$. We envisage that these observations could be useful for future electronics applications: namely design of high-current devices. Also, we speculate that a simple way to generate high-MI could be to simply achieve high current in certain systems. Strikingly, it is found to be possible to generate states with maximal MI for the quantum dot even at high temperatures by simply applying a large `source-drain' voltage. Although MI includes both quantum and classical correlation, at very low temperatures, the system gets closer and closer to pure-states, in which case MI is exactly $2$ times the entanglement entropy, thus in order to exploit our understanding for the harnessing pure quantum correlations in an experimental setting, low temperatures would be appropriate. Further studies on this would be desirable to clarify and perhaps quantify some of these dependences, and also to consider other kinds of systems to see if these features are universal.

\begin{acknowledgments}
A.S gratefully acknowledges helpful discussions with Simone Paganelli, Giacomo Gori, Pasquale Sodano, and Andrea Trombettoni. Both authors thank Tal Levy and Eli Wilner for helpful discussions. A.S is grateful to The Center for Nanoscience and Nanotechnology at Tel Aviv University for a postdoctoral fellowship. E.R thanks the Marko and Lucie Chaoul Chair. This work was supported by the Israel Science Foundation (Grant 611/11). We are grateful to the anonymous referee and Peter Young for helping tighten our understanding of the notion of mutual information in relation to entanglement, and Shivaji Sondhi for inspiring an analytical study of the high-temperature quadratic dependence, and Alexei Andreanov for a remark about the spectral function of a matrix.
\end{acknowledgments}

\bibliography{ent_landauer}

\begin{thebibliography}{29}%
\makeatletter
\providecommand \@ifxundefined [1]{%
 \@ifx{#1\undefined}
}%
\providecommand \@ifnum [1]{%
 \ifnum #1\expandafter \@firstoftwo
 \else \expandafter \@secondoftwo
 \fi
}%
\providecommand \@ifx [1]{%
 \ifx #1\expandafter \@firstoftwo
 \else \expandafter \@secondoftwo
 \fi
}%
\providecommand \natexlab [1]{#1}%
\providecommand \enquote  [1]{``#1''}%
\providecommand \bibnamefont  [1]{#1}%
\providecommand \bibfnamefont [1]{#1}%
\providecommand \citenamefont [1]{#1}%
\providecommand \href@noop [0]{\@secondoftwo}%
\providecommand \href [0]{\begingroup \@sanitize@url \@href}%
\providecommand \@href[1]{\@@startlink{#1}\@@href}%
\providecommand \@@href[1]{\endgroup#1\@@endlink}%
\providecommand \@sanitize@url [0]{\catcode `\\12\catcode `\$12\catcode
  `\&12\catcode `\#12\catcode `\^12\catcode `\_12\catcode `\%12\relax}%
\providecommand \@@startlink[1]{}%
\providecommand \@@endlink[0]{}%
\providecommand \url  [0]{\begingroup\@sanitize@url \@url }%
\providecommand \@url [1]{\endgroup\@href {#1}{\urlprefix }}%
\providecommand \urlprefix  [0]{URL }%
\providecommand \Eprint [0]{\href }%
\providecommand \doibase [0]{http://dx.doi.org/}%
\providecommand \selectlanguage [0]{\@gobble}%
\providecommand \bibinfo  [0]{\@secondoftwo}%
\providecommand \bibfield  [0]{\@secondoftwo}%
\providecommand \translation [1]{[#1]}%
\providecommand \BibitemOpen [0]{}%
\providecommand \bibitemStop [0]{}%
\providecommand \bibitemNoStop [0]{.\EOS\space}%
\providecommand \EOS [0]{\spacefactor3000\relax}%
\providecommand \BibitemShut  [1]{\csname bibitem#1\endcsname}%
\let\auto@bib@innerbib\@empty
\bibitem [{\citenamefont {Haug}\ and\ \citenamefont
  {Jauho}(2007)}]{haug2007quantum}%
  \BibitemOpen
  \bibfield  {author} {\bibinfo {author} {\bibfnamefont {H.}~\bibnamefont
  {Haug}}\ and\ \bibinfo {author} {\bibfnamefont {A.-P.}\ \bibnamefont
  {Jauho}},\ }\href@noop {} {\emph {\bibinfo {title} {Quantum kinetics in
  transport and optics of semiconductors}}},\ Vol.\ \bibinfo {volume} {123}\
  (\bibinfo  {publisher} {Springer},\ \bibinfo {year} {2007})\BibitemShut
  {NoStop}%
\bibitem [{\citenamefont {Datta}(1997)}]{datta1997electronic}%
  \BibitemOpen
  \bibfield  {author} {\bibinfo {author} {\bibfnamefont {S.}~\bibnamefont
  {Datta}},\ }\href@noop {} {\emph {\bibinfo {title} {Electronic transport in
  mesoscopic systems}}}\ (\bibinfo  {publisher} {Cambridge university press},\
  \bibinfo {year} {1997})\BibitemShut {NoStop}%
\bibitem [{\citenamefont {Landauer}(1957)}]{landauer1957spatial}%
  \BibitemOpen
  \bibfield  {author} {\bibinfo {author} {\bibfnamefont {R.}~\bibnamefont
  {Landauer}},\ }\href {\doibase 10.1147/rd.13.0223} {\bibfield  {journal}
  {\bibinfo  {journal} {IBM Journal of Research and Development}\ }\textbf
  {\bibinfo {volume} {1}},\ \bibinfo {pages} {223} (\bibinfo {year}
  {1957})}\BibitemShut {NoStop}%
\bibitem [{\citenamefont {Landauer}(1970)}]{landauer1970electrical}%
  \BibitemOpen
  \bibfield  {author} {\bibinfo {author} {\bibfnamefont {R.}~\bibnamefont
  {Landauer}},\ }\href {\doibase 10.1080/14786437008238472} {\bibfield
  {journal} {\bibinfo  {journal} {Philosophical Magazine}\ }\textbf {\bibinfo
  {volume} {21}},\ \bibinfo {pages} {863} (\bibinfo {year} {1970})}\BibitemShut
  {NoStop}%
\bibitem [{\citenamefont {B\"uttiker}(1986)}]{buttiker1986four}%
  \BibitemOpen
  \bibfield  {author} {\bibinfo {author} {\bibfnamefont {M.}~\bibnamefont
  {B\"uttiker}},\ }\href {\doibase 10.1103/PhysRevLett.57.1761} {\bibfield
  {journal} {\bibinfo  {journal} {Phys. Rev. Lett.}\ }\textbf {\bibinfo
  {volume} {57}},\ \bibinfo {pages} {1761} (\bibinfo {year}
  {1986})}\BibitemShut {NoStop}%
\bibitem [{\citenamefont {Stefanucci}\ and\ \citenamefont {van
  Leeuwen}(2013)}]{stefanucci_nonequilibrium_2013}%
  \BibitemOpen
  \bibfield  {author} {\bibinfo {author} {\bibfnamefont {G.}~\bibnamefont
  {Stefanucci}}\ and\ \bibinfo {author} {\bibfnamefont {R.}~\bibnamefont {van
  Leeuwen}},\ }\href@noop {} {\emph {\bibinfo {title} {Nonequilibrium Many-Body
  Theory of Quantum Systems: A Modern Introduction}}}\ (\bibinfo  {publisher}
  {Cambridge University Press},\ \bibinfo {year} {2013})\BibitemShut {NoStop}%
\bibitem [{\citenamefont {M\"uhlbacher}\ and\ \citenamefont
  {Rabani}(2008)}]{lothar2008real}%
  \BibitemOpen
  \bibfield  {author} {\bibinfo {author} {\bibfnamefont {L.}~\bibnamefont
  {M\"uhlbacher}}\ and\ \bibinfo {author} {\bibfnamefont {E.}~\bibnamefont
  {Rabani}},\ }\href {\doibase 10.1103/PhysRevLett.100.176403} {\bibfield
  {journal} {\bibinfo  {journal} {Phys. Rev. Lett.}\ }\textbf {\bibinfo
  {volume} {100}},\ \bibinfo {pages} {176403} (\bibinfo {year}
  {2008})}\BibitemShut {NoStop}%
\bibitem [{\citenamefont {Weiss}\ \emph {et~al.}(2008)\citenamefont {Weiss},
  \citenamefont {Eckel}, \citenamefont {Thorwart},\ and\ \citenamefont
  {Egger}}]{weiss_iterative_2008}%
  \BibitemOpen
  \bibfield  {author} {\bibinfo {author} {\bibfnamefont {S.}~\bibnamefont
  {Weiss}}, \bibinfo {author} {\bibfnamefont {J.}~\bibnamefont {Eckel}},
  \bibinfo {author} {\bibfnamefont {M.}~\bibnamefont {Thorwart}}, \ and\
  \bibinfo {author} {\bibfnamefont {R.}~\bibnamefont {Egger}},\ }\href
  {\doibase 10.1103/PhysRevB.77.195316} {\bibfield  {journal} {\bibinfo
  {journal} {Phys. Rev. B}\ }\textbf {\bibinfo {volume} {77}},\ \bibinfo
  {pages} {195316} (\bibinfo {year} {2008})}\BibitemShut {NoStop}%
\bibitem [{\citenamefont {Werner}\ \emph {et~al.}(2009)\citenamefont {Werner},
  \citenamefont {Oka},\ and\ \citenamefont
  {Millis}}]{werner_diagrammatic_2009}%
  \BibitemOpen
  \bibfield  {author} {\bibinfo {author} {\bibfnamefont {P.}~\bibnamefont
  {Werner}}, \bibinfo {author} {\bibfnamefont {T.}~\bibnamefont {Oka}}, \ and\
  \bibinfo {author} {\bibfnamefont {A.~J.}\ \bibnamefont {Millis}},\ }\href
  {\doibase 10.1103/PhysRevB.79.035320} {\bibfield  {journal} {\bibinfo
  {journal} {Phys. Rev. B}\ }\textbf {\bibinfo {volume} {79}},\ \bibinfo
  {pages} {035320} (\bibinfo {year} {2009})}\BibitemShut {NoStop}%
\bibitem [{\citenamefont {Wang}\ and\ \citenamefont {Thoss}(2009)}]{Wang2009}%
  \BibitemOpen
  \bibfield  {author} {\bibinfo {author} {\bibfnamefont {H.}~\bibnamefont
  {Wang}}\ and\ \bibinfo {author} {\bibfnamefont {M.}~\bibnamefont {Thoss}},\
  }\href@noop {} {\bibfield  {journal} {\bibinfo  {journal} {J. Chem. Phys.}\
  }\textbf {\bibinfo {volume} {131}},\ \bibinfo {pages} {024114} (\bibinfo
  {year} {2009})}\BibitemShut {NoStop}%
\bibitem [{\citenamefont {Gull}\ \emph {et~al.}(2010)\citenamefont {Gull},
  \citenamefont {Reichman},\ and\ \citenamefont {Millis}}]{gull2010bold}%
  \BibitemOpen
  \bibfield  {author} {\bibinfo {author} {\bibfnamefont {E.}~\bibnamefont
  {Gull}}, \bibinfo {author} {\bibfnamefont {D.~R.}\ \bibnamefont {Reichman}},
  \ and\ \bibinfo {author} {\bibfnamefont {A.~J.}\ \bibnamefont {Millis}},\
  }\href {\doibase 10.1103/PhysRevB.82.075109} {\bibfield  {journal} {\bibinfo
  {journal} {Phys. Rev. B}\ }\textbf {\bibinfo {volume} {82}},\ \bibinfo
  {pages} {075109} (\bibinfo {year} {2010})}\BibitemShut {NoStop}%
\bibitem [{\citenamefont {Segal}\ \emph {et~al.}(2010)\citenamefont {Segal},
  \citenamefont {Millis},\ and\ \citenamefont {Reichman}}]{Segal10}%
  \BibitemOpen
  \bibfield  {author} {\bibinfo {author} {\bibfnamefont {D.}~\bibnamefont
  {Segal}}, \bibinfo {author} {\bibfnamefont {A.~J.}\ \bibnamefont {Millis}}, \
  and\ \bibinfo {author} {\bibfnamefont {D.~R.}\ \bibnamefont {Reichman}},\
  }\href@noop {} {\bibfield  {journal} {\bibinfo  {journal} {Phys. Rev. B}\
  }\textbf {\bibinfo {volume} {82}},\ \bibinfo {pages} {205323} (\bibinfo
  {year} {2010})}\BibitemShut {NoStop}%
\bibitem [{\citenamefont {{H\"utzen}}\ \emph {et~al.}(2012)\citenamefont
  {{H\"utzen}}, \citenamefont {Weiss}, \citenamefont {Thorwart},\ and\
  \citenamefont {Egger}}]{Huetzen12}%
  \BibitemOpen
  \bibfield  {author} {\bibinfo {author} {\bibfnamefont {R.}~\bibnamefont
  {{H\"utzen}}}, \bibinfo {author} {\bibfnamefont {S.}~\bibnamefont {Weiss}},
  \bibinfo {author} {\bibfnamefont {M.}~\bibnamefont {Thorwart}}, \ and\
  \bibinfo {author} {\bibfnamefont {R.}~\bibnamefont {Egger}},\ }\href@noop {}
  {\bibfield  {journal} {\bibinfo  {journal} {Phys. Rev. B}\ }\textbf {\bibinfo
  {volume} {85}},\ \bibinfo {pages} {121408(R)} (\bibinfo {year}
  {2012})}\BibitemShut {NoStop}%
\bibitem [{\citenamefont {Simine}\ and\ \citenamefont
  {Segal}(2013)}]{Simine13}%
  \BibitemOpen
  \bibfield  {author} {\bibinfo {author} {\bibfnamefont {L.}~\bibnamefont
  {Simine}}\ and\ \bibinfo {author} {\bibfnamefont {D.}~\bibnamefont {Segal}},\
  }\href@noop {} {\bibfield  {journal} {\bibinfo  {journal} {J. Comp. Phys.}\
  }\textbf {\bibinfo {volume} {138}},\ \bibinfo {pages} {214111} (\bibinfo
  {year} {2013})}\BibitemShut {NoStop}%
\bibitem [{\citenamefont {Amico}\ \emph {et~al.}(2008)\citenamefont {Amico},
  \citenamefont {Fazio}, \citenamefont {Osterloh},\ and\ \citenamefont
  {Vedral}}]{amico2008entanglement}%
  \BibitemOpen
  \bibfield  {author} {\bibinfo {author} {\bibfnamefont {L.}~\bibnamefont
  {Amico}}, \bibinfo {author} {\bibfnamefont {R.}~\bibnamefont {Fazio}},
  \bibinfo {author} {\bibfnamefont {A.}~\bibnamefont {Osterloh}}, \ and\
  \bibinfo {author} {\bibfnamefont {V.}~\bibnamefont {Vedral}},\ }\href
  {\doibase 10.1103/RevModPhys.80.517} {\bibfield  {journal} {\bibinfo
  {journal} {Rev. Mod. Phys.}\ }\textbf {\bibinfo {volume} {80}},\ \bibinfo
  {pages} {517} (\bibinfo {year} {2008})}\BibitemShut {NoStop}%
\bibitem [{\citenamefont {\ifmmode \check{Z}\else
  \v{Z}\fi{}nidari\ifmmode~\check{c}\else \v{c}\fi{}}(2012)}]{znidaric2012}%
  \BibitemOpen
  \bibfield  {author} {\bibinfo {author} {\bibfnamefont {M.}~\bibnamefont
  {\ifmmode \check{Z}\else \v{Z}\fi{}nidari\ifmmode~\check{c}\else
  \v{c}\fi{}}},\ }\href {\doibase 10.1103/PhysRevA.85.012324} {\bibfield
  {journal} {\bibinfo  {journal} {Phys. Rev. A}\ }\textbf {\bibinfo {volume}
  {85}},\ \bibinfo {pages} {012324} (\bibinfo {year} {2012})}\BibitemShut
  {NoStop}%
\bibitem [{\citenamefont {Eisler}\ and\ \citenamefont
  {Zimbor\'as}(2014)}]{eisler2014}%
  \BibitemOpen
  \bibfield  {author} {\bibinfo {author} {\bibfnamefont {V.}~\bibnamefont
  {Eisler}}\ and\ \bibinfo {author} {\bibfnamefont {Z.}~\bibnamefont
  {Zimbor\'as}},\ }\href {\doibase 10.1103/PhysRevA.89.032321} {\bibfield
  {journal} {\bibinfo  {journal} {Phys. Rev. A}\ }\textbf {\bibinfo {volume}
  {89}},\ \bibinfo {pages} {032321} (\bibinfo {year} {2014})}\BibitemShut
  {NoStop}%
\bibitem [{\citenamefont {\ifmmode \check{Z}\else
  \v{Z}\fi{}nidari\ifmmode~\check{c}\else \v{c}\fi{}}\ \emph
  {et~al.}(2008)\citenamefont {\ifmmode \check{Z}\else
  \v{Z}\fi{}nidari\ifmmode~\check{c}\else \v{c}\fi{}}, \citenamefont {Prosen},\
  and\ \citenamefont {Pi\ifmmode~\check{z}\else \v{z}\fi{}orn}}]{znidaric2008}%
  \BibitemOpen
  \bibfield  {author} {\bibinfo {author} {\bibfnamefont {M.}~\bibnamefont
  {\ifmmode \check{Z}\else \v{Z}\fi{}nidari\ifmmode~\check{c}\else
  \v{c}\fi{}}}, \bibinfo {author} {\bibfnamefont {T.~c.~v.}\ \bibnamefont
  {Prosen}}, \ and\ \bibinfo {author} {\bibfnamefont {I.}~\bibnamefont
  {Pi\ifmmode~\check{z}\else \v{z}\fi{}orn}},\ }\href {\doibase
  10.1103/PhysRevA.78.022103} {\bibfield  {journal} {\bibinfo  {journal} {Phys.
  Rev. A}\ }\textbf {\bibinfo {volume} {78}},\ \bibinfo {pages} {022103}
  (\bibinfo {year} {2008})}\BibitemShut {NoStop}%
\bibitem [{\citenamefont {Plenio}\ and\ \citenamefont
  {Virmani}(2005)}]{plenio2005introduction}%
  \BibitemOpen
  \bibfield  {author} {\bibinfo {author} {\bibfnamefont {M.~B.}\ \bibnamefont
  {Plenio}}\ and\ \bibinfo {author} {\bibfnamefont {S.}~\bibnamefont
  {Virmani}},\ }\href@noop {} {\bibfield  {journal} {\bibinfo  {journal} {arXiv
  preprint quant-ph/0504163}\ } (\bibinfo {year} {2005})}\BibitemShut {NoStop}%
\bibitem [{\citenamefont {Schollw\"ock}(2005)}]{schollwock2005density}%
  \BibitemOpen
  \bibfield  {author} {\bibinfo {author} {\bibfnamefont {U.}~\bibnamefont
  {Schollw\"ock}},\ }\href {\doibase 10.1103/RevModPhys.77.259} {\bibfield
  {journal} {\bibinfo  {journal} {Rev. Mod. Phys.}\ }\textbf {\bibinfo {volume}
  {77}},\ \bibinfo {pages} {259} (\bibinfo {year} {2005})}\BibitemShut
  {NoStop}%
\bibitem [{\citenamefont {Klich}\ and\ \citenamefont
  {Levitov}(2009)}]{klich2009quantum}%
  \BibitemOpen
  \bibfield  {author} {\bibinfo {author} {\bibfnamefont {I.}~\bibnamefont
  {Klich}}\ and\ \bibinfo {author} {\bibfnamefont {L.}~\bibnamefont
  {Levitov}},\ }\href {\doibase 10.1103/PhysRevLett.102.100502} {\bibfield
  {journal} {\bibinfo  {journal} {Phys. Rev. Lett.}\ }\textbf {\bibinfo
  {volume} {102}},\ \bibinfo {pages} {100502} (\bibinfo {year}
  {2009})}\BibitemShut {NoStop}%
\bibitem [{\citenamefont {Chien}\ \emph {et~al.}(2014)\citenamefont {Chien},
  \citenamefont {Di~Ventra},\ and\ \citenamefont {Zwolak}}]{chien2014landauer}%
  \BibitemOpen
  \bibfield  {author} {\bibinfo {author} {\bibfnamefont {C.-C.}\ \bibnamefont
  {Chien}}, \bibinfo {author} {\bibfnamefont {M.}~\bibnamefont {Di~Ventra}}, \
  and\ \bibinfo {author} {\bibfnamefont {M.}~\bibnamefont {Zwolak}},\ }\href
  {\doibase 10.1103/PhysRevA.90.023624} {\bibfield  {journal} {\bibinfo
  {journal} {Phys. Rev. A}\ }\textbf {\bibinfo {volume} {90}},\ \bibinfo
  {pages} {023624} (\bibinfo {year} {2014})}\BibitemShut {NoStop}%
\bibitem [{\citenamefont {Dhar}\ \emph {et~al.}(2012)\citenamefont {Dhar},
  \citenamefont {Saito},\ and\ \citenamefont
  {H\"anggi}}]{dhar2012nonequilibrium}%
  \BibitemOpen
  \bibfield  {author} {\bibinfo {author} {\bibfnamefont {A.}~\bibnamefont
  {Dhar}}, \bibinfo {author} {\bibfnamefont {K.}~\bibnamefont {Saito}}, \ and\
  \bibinfo {author} {\bibfnamefont {P.}~\bibnamefont {H\"anggi}},\ }\href
  {\doibase 10.1103/PhysRevE.85.011126} {\bibfield  {journal} {\bibinfo
  {journal} {Phys. Rev. E}\ }\textbf {\bibinfo {volume} {85}},\ \bibinfo
  {pages} {011126} (\bibinfo {year} {2012})}\BibitemShut {NoStop}%
\bibitem [{\citenamefont {Hershfield}(1993)}]{hershfield1993reformulation}%
  \BibitemOpen
  \bibfield  {author} {\bibinfo {author} {\bibfnamefont {S.}~\bibnamefont
  {Hershfield}},\ }\href {\doibase 10.1103/PhysRevLett.70.2134} {\bibfield
  {journal} {\bibinfo  {journal} {Phys. Rev. Lett.}\ }\textbf {\bibinfo
  {volume} {70}},\ \bibinfo {pages} {2134} (\bibinfo {year}
  {1993})}\BibitemShut {NoStop}%
\bibitem [{\citenamefont {Peschel}(2003)}]{peschel2003letter}%
  \BibitemOpen
  \bibfield  {author} {\bibinfo {author} {\bibfnamefont {I.}~\bibnamefont
  {Peschel}},\ }\href@noop {} {\bibfield  {journal} {\bibinfo  {journal} {J.
  Phys. A}\ }\textbf {\bibinfo {volume} {36}},\ \bibinfo {pages} {L205}
  (\bibinfo {year} {2003})}\BibitemShut {NoStop}%
\bibitem [{\citenamefont {Henderson}\ and\ \citenamefont
  {Vedral}(2001)}]{henderson2001classical}%
  \BibitemOpen
  \bibfield  {author} {\bibinfo {author} {\bibfnamefont {L.}~\bibnamefont
  {Henderson}}\ and\ \bibinfo {author} {\bibfnamefont {V.}~\bibnamefont
  {Vedral}},\ }\href@noop {} {\bibfield  {journal} {\bibinfo  {journal} {J
  Phys. A}\ }\textbf {\bibinfo {volume} {34}},\ \bibinfo {pages} {6899}
  (\bibinfo {year} {2001})}\BibitemShut {NoStop}%
\bibitem [{\citenamefont {Wolf}\ \emph {et~al.}(2008)\citenamefont {Wolf},
  \citenamefont {Verstraete}, \citenamefont {Hastings},\ and\ \citenamefont
  {Cirac}}]{wolf2008area}%
  \BibitemOpen
  \bibfield  {author} {\bibinfo {author} {\bibfnamefont {M.~M.}\ \bibnamefont
  {Wolf}}, \bibinfo {author} {\bibfnamefont {F.}~\bibnamefont {Verstraete}},
  \bibinfo {author} {\bibfnamefont {M.~B.}\ \bibnamefont {Hastings}}, \ and\
  \bibinfo {author} {\bibfnamefont {J.~I.}\ \bibnamefont {Cirac}},\ }\href@noop
  {} {\bibfield  {journal} {\bibinfo  {journal} {Phys.Rev.Lett.}\ }\textbf
  {\bibinfo {volume} {100}},\ \bibinfo {pages} {070502} (\bibinfo {year}
  {2008})}\BibitemShut {NoStop}%
\bibitem [{\citenamefont {Swenson}\ \emph {et~al.}(2011)\citenamefont
  {Swenson}, \citenamefont {Levy}, \citenamefont {Cohen}, \citenamefont
  {Rabani},\ and\ \citenamefont {Miller}}]{swenson2011application}%
  \BibitemOpen
  \bibfield  {author} {\bibinfo {author} {\bibfnamefont {D.~W.~H.}\
  \bibnamefont {Swenson}}, \bibinfo {author} {\bibfnamefont {T.}~\bibnamefont
  {Levy}}, \bibinfo {author} {\bibfnamefont {G.}~\bibnamefont {Cohen}},
  \bibinfo {author} {\bibfnamefont {E.}~\bibnamefont {Rabani}}, \ and\ \bibinfo
  {author} {\bibfnamefont {W.~H.}\ \bibnamefont {Miller}},\ }\href {\doibase
  http://dx.doi.org/10.1063/1.3583366} {\bibfield  {journal} {\bibinfo
  {journal} {The Journal of Chemical Physics}\ }\textbf {\bibinfo {volume}
  {134}},\ \bibinfo {eid} {164103} (\bibinfo {year} {2011})}\BibitemShut
  {NoStop}%
\bibitem [{\citenamefont {Peschel}(2012)}]{peschel2012special}%
  \BibitemOpen
  \bibfield  {author} {\bibinfo {author} {\bibfnamefont {I.}~\bibnamefont
  {Peschel}},\ }\href {\doibase 10.1007/s13538-012-0074-1} {\bibfield
  {journal} {\bibinfo  {journal} {Brazilian Journal of Physics}\ }\textbf
  {\bibinfo {volume} {42}},\ \bibinfo {pages} {267} (\bibinfo {year}
  {2012})}\BibitemShut {NoStop}%
\end{thebibliography}%

\appendix
\section{Current}
\label{sec:current}
In terms of the fermionic operators
\begin{equation}
\label{eqn:atoc}
a_{\alpha} = \sum_{i=1}^{N}\psi_{\alpha}(i)c_{i},
\end{equation}
the Hamiltonian becomes $\mathcal{H}=\sum_{\alpha=1}^{N}e_{\alpha}a^{\dagger}_{\alpha}a_{\alpha}$.

Since:
\begin{align}
\sum_{\alpha=1}^{N}\psi_{\alpha}^{*}(j)a_{\alpha} &= \sum_{i=1}^{N}\sum_{\alpha=1}^{N}\psi_{\alpha}^{*}(j)\psi_{\alpha}(i)c_{i}\\
                                              &= \sum_{i=1}^{N}\delta_{i,j}c_{i},\nonumber
\end{align}
\begin{align}
\label{eqn:ctoa}
c_{j} = \sum_{\alpha=1}^{N}\psi_{\alpha}^{*}(j)a_{\alpha}.
\end{align}

We note that for $j = 1, \cdots, N_{L}$ these destruction operators refer to the left lead,  for $j=(N_{L}+1),\cdots, (N_{L}+N_{R})$ refer to the right lead, and for $j=N$ to the dot ($c_{N}\equiv d$).

The time evolution of the Heisenberg operators $a_{\alpha}(t)$ can be easily written down.
\begin{align}
\dot{a}_{\alpha} &= \frac{1}{i\hbar}\Big[a_{\alpha},\sum_{\beta}e_{\beta}a_{\beta}^{\dagger}a_{\beta}\Big] \\
               &=  \frac{1}{i\hbar}e_{\alpha}a_{\alpha}. \nonumber
\end{align}
Therefore (with $\hbar=1$), 
\begin{align}
\label{eqn:aoft}
a_{\alpha}(t) = e^{-ite_{\alpha}}a_{\alpha}(0).
\end{align}
Thus, with the help of the relation between the $c$ and $a$ operators, the time evolution of all the original operators $c_{j}(t)$ is easily obtained.

The left current is given by the change in occupancy of the left electrode:
\begin{align}
I_{L}(t) &= -e \frac{d}{dt}\Big\langle\sum_{j\in L}c_{j}^{\dagger}c_{j}\Big\rangle\\
        &= -\frac{ei}{\hbar}\Big\langle\sum_{j\in L}t_{j}(d^{\dagger}c_{j}-c_{j}^{\dagger}d)\Big\rangle.
\end{align}

From Eqn.~\ref{eqn:aoft}, and from the relation between the $c$ and $a$ operators, we can write
\begin{align}
\Big\langle c_{p}^{\dagger}(t)c_{q}(t)\Big\rangle  &=  \sum_{\alpha\beta}\psi_{\alpha}(p)\psi_{\beta}^{*}(q)\big\langle a_{\alpha}^{\dagger}(t)a_{\beta}(t)\big\rangle \\
                                                &= \sum_{\alpha\beta}\psi_{\alpha}(p)\psi_{\beta}^{*}(q)e^{it(e_{\alpha}-e_{\beta})}\big\langle a_{\alpha}^{\dagger}(0)a_{\beta}(0)\big\rangle \nonumber\\
                                                &\hspace{-3em}= \sum_{\alpha\beta}e^{it(e_{\alpha}-e_{\beta})}\psi_{\alpha}(p)\psi_{\beta}^{*}(q)\sum_{kl}\psi_{\alpha}^{*}(k)\psi_{\beta}(l)\big\langle c_{k}^{\dagger}(0)c_{l}(0)\big\rangle \nonumber
\end{align}
The left current can now be written as (with $\hbar=1$,$e=1$):
\begin{widetext}
\begin{align}
I_{L}(t) 
        &= 2\sum_{j\in L}t_{j}\sum_{\alpha\beta}\sum_{kl}Im[e^{it(e_{\alpha}-e_{\beta})}\psi_{\alpha}(N)\psi_{\beta}^{*}(j)\psi_{\alpha}^{*}(k)\psi_{\beta}(l)]\big\langle c_{k}^{\dagger}(0)c_{l}(0)\big\rangle 
\end{align}
\end{widetext}
where we have used the fact that the dot operator corresponds to the $N^{th}$ mode in our representation. A similar expression exists for the right current. Therefore combining the two and defining 
\begin{align}
\tilde{t}_{j} &= t_{j}  \quad\quad\quad\quad\quad\quad\text{if}\quad j\in L  \nonumber\\
             &= -t_{j} \quad\quad\quad\quad\quad\;\text{if}\quad j\in R  \nonumber,
\end{align}
we can compute the overall current 
\begin{widetext}
\begin{align}
I(t) 
        &= \sum_{j}\tilde{t}_{j}\sum_{\alpha\beta}\sum_{kl}Im[e^{it(e_{\alpha}-e_{\beta})}\psi_{\alpha}(N)\psi_{\beta}^{*}(j)\psi_{\alpha}^{*}(k)\psi_{\beta}(l)]\big\langle c_{k}^{\dagger}(0)c_{l}(0)\big\rangle. 
\end{align}
\end{widetext}
Our initial condition is such that 
\begin{align}
\langle c_{k}^{\dagger}(0)c_{l}(0)\rangle = \delta_{kl}f
\end{align}
where $f$ is the Fermi function $f(\epsilon_{k}-\mu_{L})$ for modes on the left lead, the Fermi function $f(\epsilon_{k}-\mu_{R})$ for modes on the right lead, and the occupancy $n_{0}$ for the dot mode. Thus the summation simplifies for the expression for current.

\section{Determination of the $H$ matrix in terms of the correlator matrix $C$}
\label{sec:HC}
Because of our choice of initial density matrix which is of the form $\rho(0) = \frac{e^{-\sum_{i,j}H_{i,j}(0)c_{i}^{\dagger}c_{j}}}{Z}$, Wick's theorem holds at $t=0$. Therefore 
\begin{widetext}
\begin{align}
\big\langle c_{n}^{\dagger}(t)c_{m}^{\dagger}(t)c_{l}(t)c_{j}(t)\big\rangle &= \sum_{\alpha\beta\gamma\delta}\psi_{\alpha}(n)\psi_{\beta}(m)\psi_{\gamma}^{*}(l)\psi_{\delta}^{*}(j)\big\langle a_{\alpha}^{\dagger}(t)a_{\beta}^{\dagger}(t)a_{\gamma}(t)a_{\delta}(t)\big\rangle \\
                  &=\sum_{\alpha\beta\gamma\delta}\psi_{\alpha}(n)\psi_{\beta}(m)\psi_{\gamma}^{*}(l)\psi_{\delta}^{*}(j)e^{it(e_{\alpha}+e_{\beta}-e_{\gamma}-e_{\delta})}\big\langle a_{\alpha}^{\dagger}(0)a_{\beta}^{\dagger}(0)a_{\gamma}(0)a_{\delta}(0)\big\rangle \nonumber\\
                  &=\sum_{\alpha\beta\gamma\delta}\psi_{\alpha}(n)\psi_{\beta}(m)\psi_{\gamma}^{*}(l)\psi_{\delta}^{*}(j)e^{it(e_{\alpha}+e_{\beta}-e_{\gamma}-e_{\delta})}\nonumber\\
                  &\quad\quad\quad\quad\Big[\big\langle a_{\alpha}^{\dagger}(0)a_{\delta}(0)\big\rangle\big\langle a_{\beta}^{\dagger}(0)a_{\gamma}(0)\big\rangle-\big\langle a_{\alpha}^{\dagger}(0)a_{\gamma}(0)\big\rangle\big\langle a_{\beta}^{\dagger}(0)a_{\delta}(0)\big\rangle\Big]\nonumber\\
                  &=\big\langle c_{n}^{\dagger}(t)c_{j}(t)\big\rangle\big\langle c_{m}^{\dagger}(t)c_{l}(t)\big\rangle-\big\langle c_{n}^{\dagger}(t)c_{l}(t)\big\rangle\big\langle c_{m}^{\dagger}(t)c_{j}(t)\big\rangle\nonumber,
\end{align}
\end{widetext}
thus showing that Wick's theorem holds at arbitrary time. Therefore, we posit that $\rho(t)$ has the form 
\begin{equation}
\rho(t) = \frac{e^{-\sum_{i,j}H_{i,j}(t)c_{i}^{\dagger}c_{j}}}{Z},
\end{equation}
and now calculate the matrix $H_{i,j}(t)$. For clarity, we make the time dependence of quantities implicit henceforth.
Let $\phi_{\sigma}(i)$ be the eigenfunctions of $H$ with eigenvalues $h_{\sigma}$. Then the transformation to new fermion operators $a_{\sigma}$ 
\begin{equation}
a_{\sigma} = \sum_{i=1}^{L}\phi_{\sigma}(i)c_{i}
\end{equation}
gives
\begin{equation}
\label{eqn:densitydiag}
\rho(t) = \frac{\exp\Big(-\sum_{\sigma=1}^{L}h_{\sigma}a_{\sigma}^{\dagger}a_{\sigma}\Big)}{\prod_{\sigma=1}^{L}(1+\exp(-h_{\sigma}))}.
\end{equation}
It turns out that the correlator matrix defined as
\begin{equation}
\label{eqn:C}
C_{ij} = \Tr(\rho(t) c_{i}^{\dagger}c_{j}),
\end{equation}
is also diagonalized by precisely the same eigenfunctions that diagonalize the matrix $H$. This is seen simply as follows:
\begin{align}
\sum_{j}C_{ij}\phi_{\sigma}(j) &= \Tr(c_{i}^{\dagger}\sum_{j}c_{j}\phi_{\sigma}(j)\rho(t))\\
                           &= \Tr(c_{i}a_{\alpha}\rho(t))\nonumber\\
                           &= \Tr(\sum_{\beta}\phi_{\beta}(i)a_{\beta}^{\dagger}a_{\alpha}\rho(t))\nonumber\\
                           &= \Tr(a_{\alpha}^{\dagger}a_{\alpha}\rho(t))\phi_{\sigma}(j)\nonumber\\
                           &= \frac{e^{-h_{\alpha}}}{1+e^{-h_{\alpha}}}\phi_{\sigma}(j)\nonumber,
\end{align}
where we have used the precise diagonal form of Eqn.~\ref{eqn:densitydiag} to simplify. Thus the matrices $C$ and $H$ share eigenfunctions, and the relationship between their eigenvalues implies that we can write
\begin{equation}
\label{eqn:HandC}
\exp(H) = (-C+1)C^{-1},
\end{equation}
which completes the procedure for how to obtain the full nonequilibrium density matrix as an effective thermal density matrix.

\section{Von-Neumann Entropy of a subsystem of the full non-equilibrium mixed state density matrix}
\label{sec:vN}
The reduced density matrices themselves can be written as thermal density matrices of the form $\rho_{red} = \frac{e^{-\sum_{i,j}H_{i,j}c_{i}^{\dagger}c_{j}}}{Z}$ by the same argument given in the previous section: Wick's theorem holds within the reduced subspace, and the reduced density matrix is unique, and Wick's theorem holds for a thermal type of density matrix. Once again the $H$ matrix is related to the correlator matrix $C$ by the formula $\exp(H) = (-C+1)C^{-1}$, with the crucial difference that the correlator matrix indices run only over the reduced subspace. We now obtain a simple formula for the von-Neumann entropy of any subspace $\mathcal{G}$ 
\begin{equation}
S_{\mathcal{G}} = -\Tr(\rho_{\mathcal{G}}\ln{\rho_{\mathcal{G}}}),
\end{equation}
\begin{align}
S_{\mathcal{G}} &= -\Tr\Big[\frac{1}{Z}\exp\big(-\sum_{\sigma=1}^{L}h_{\sigma}a_{\sigma}^{\dagger}a_{\sigma}\big)\Big(\ln(\frac{1}{Z})-\sum_{\sigma=1}^{L}h_{\sigma}a_{\sigma}^{\dagger}a_{\sigma}\Big)\Big]\\
             &= \sum_{\sigma=1}^{L}\ln\big(1+\exp(-h_{\sigma})\big)+\sum_{\sigma=1}^{L}\Big[h_{\sigma}\exp(-h_{\sigma})\frac{1}{1+\exp(-h_{\sigma})}\Big]\notag\\
             &= \sum_{\sigma=1}^{L}\Big[\ln\big(1+\exp(-h_{\sigma})\big)+\frac{h_{\sigma}}{1+\exp(h_{\sigma})}\Big]\notag.
\end{align}
But, we know that 
\begin{equation}
\exp(h_{\sigma}) = \frac{1-C_{\sigma}}{C_{\sigma}},
\end{equation}
where $C_{\sigma}$ is the corresponding eigenvalue of the correlator matrix in Eqn.~\ref{eqn:C} defined with the subspace $\mathcal{G}$. Therefore,
\begin{align}
S_{\mathcal{G}} &= \sum_{\sigma=1}^{L}\Big[-(1-C_{\sigma})\ln(1-C_{\sigma})-C_{\sigma}\ln C_{\sigma}\Big].
\end{align}

\section{Quantum fluctuations in Current}
\label{sec:N}
Quantum noise is defined as
\begin{equation}
\overline{N} = \frac{1}{2} \langle\{(\hat{I}(t)-I(t)),(\hat{I}(t_{0})-I(t_{0}))\}\rangle
\end{equation}
or equivalently,
\begin{equation}
\overline{N} = \frac{1}{2} \langle\{\hat{I}(t),\hat{I}(t_{0})\}\rangle - I(t)I(t_{0}).
\end{equation}

We recall that the current operator can be written as
\begin{align}
\hat{I}(t) &= -\frac{ei}{2\hbar}\Big[\sum_{j\in L,R}\tilde{t}_{j}\Big(d^{\dagger}(t)c_{j}(t)-c_{j}^{\dagger}(t)d(t)\Big)  \Big], \quad\quad\text{where} \\
\tilde{t}_{j} &= t_{j}  \quad\quad\quad\quad\quad\quad\text{if}\quad j\in L  \nonumber\\
             &= -t_{j} \quad\quad\quad\quad\quad\;\text{if}\quad j\in R.  \nonumber
\end{align}
Therefore, quantum noise operator is given by:
\begin{widetext}
\begin{align}
\hat{N} = -\frac{e^{2}}{8\hbar^{2}}\sum_{i,j}\hat{t}_{i}\hat{t}_{j}\Bigg[ &d^{\dagger}(t)c_{i}(t)d^{\dagger}(t_{0})c_{j}(t_{0})-c_{i}^{\dagger}(t)d(t)d^{\dagger}(t_{0})c_{j}(t_{0})\\
&-d^{\dagger}(t)c_{i}(t)c_{j}^{\dagger}(t_{0})d(t_{0})+c_{i}(t)c(t)c_{j}^{\dagger}(t_{0})d(t_{0}) \nonumber\\  
&+d^{\dagger}(t_{0})c_{j}(t_{0})d^{\dagger}(t)c_{i}(t)-d^{\dagger}(t_{0})c_{j}(t_{0})c_{i}^{\dagger}(t)d(t)\nonumber\\
&-c_{j}^{\dagger}(t_{0})d(t_{0})d^{\dagger}(t)c_{i}(t)+c_{j}^{\dagger}(t_{0})d(t_{0})c_{i}^{\dagger}(t)d(t)\Bigg] - I(t)I(t_{0}).\nonumber 
\end{align}
\end{widetext}
The time dependence of the operators $c_{i}(t)$ can all be written down in terms of the eigenvalues $e_{\alpha}$ and wavefunctions $\psi_{\alpha}(i)$ of the hopping matrix $T_{ij}$.
Assuming that the wavefunctions $\psi_{\alpha}(i)$ of the hopping matrix are all real (they can be made to be real since the hopping matrix is real and symmetric), we can, after a long, tedious calculation, write down a cute, compact expression for quantum noise:

\begin{align}
\overline{N} = -\frac{e^{2}}{4\hbar^{2}}\sum_{k,l}\operatorname{Re}{[\mtx{U}_{kl}(t)\mtx{V}_{kl}(t_{0})]} - I(t)I(t_{0}),
\end{align} 
where the complex matrices $\mtx{U},\mtx{V}$ are defined as follows:
\begin{widetext}
\begin{align}
\mtx{U}_{kl}(t) &= \mtx{W}_{Nk}(t)\sum_{j}\tilde{t}_{j}\mtx{W}_{jl}^{*}(t)\\
\mtx{V}_{kl}(t_{0}) &= (f_{k}+f_{l}-2f_{k}f_{l})[\mtx{W}_{Nl}(t_{0})\sum_{j}\tilde{t}_{j}\mtx{W}_{jk}^{*}(t_{0})-\mtx{W}_{Nk}^{*}(t_{0})\sum_{j}\tilde{t}_{j}\mtx{W}_{jl}(t_{0})]\\
                  &\quad\quad\quad + i 4\delta_{kl}f_{k}\operatorname{Im}[\sum_{jm}\tilde{t}_{j}f_{m}\mtx{W}_{Nm}(t_{0})\mtx{W}_{jm}^{*}(t_{0})]\nonumber,
\end{align}
\end{widetext}
where the matrix $\mtx{W}(t)$ is defined as
\begin{align}
\mtx{W}_{mn}(t) = \sum_{\alpha}e^{i t e_{\alpha}}\psi_{\alpha}(m)\psi_{\alpha}(n),
\end{align}
and $f_{k}, k=1,\cdots,(N-1)$ are the Fermi-Dirac distribution functions for the left and right leads and $f_{N}=n_{0}$ the initial dot population, which together define the initial density matrix of our system. In the above equations, the indices $k,l,m,n$ can all take values from $1,\cdots,N$, i.e the left and right leads and the dot, the label $N$ referring to the quantum dot.

When we consider the above quantum noise at a particular instance of time, it is just the quantum fluctuation in current,
\begin{equation}
\overline{N} = \langle\hat{I}(t)^{2}\rangle - I(t)^{2},
\end{equation}
which can again be computed by the methods described above.

\end{document}